\documentclass[twoside,twocolumn]{article}

\usepackage{blindtext} % Package to generate dummy text throughout this template 

\usepackage[sc]{mathpazo} % Use the Palatino font
\usepackage[T1]{fontenc} % Use 8-bit encoding that has 256 glyphs
\usepackage{microtype} % Slightly tweak font spacing for aesthetics

\usepackage[english]{babel} % Language hyphenation and typographical rules

\usepackage[hmarginratio=1:1,top=32mm,columnsep=20pt]{geometry} % Document margins
\usepackage[hang, small,labelfont=bf,up,textfont=it,up]{caption} % Custom captions under/above floats in tables or figures
\usepackage{booktabs} % Horizontal rules in tables

\usepackage{lettrine} % The lettrine is the first enlarged letter at the beginning of the text

\usepackage{enumitem} % Customized lists
\usepackage{graphicx}
\DeclareGraphicsExtensions{.pdf,.png,.jpg}
\usepackage{enotez}
\setenotez{list-name={References}}
\usepackage{hyperref}

\setlist[itemize]{noitemsep} % Make itemize lists more compact

\usepackage{abstract} % Allows abstract customization
\usepackage{titlesec} % Allows customization of titles
\renewcommand\thesection{\Roman{section}} % Roman numerals for the sections
\renewcommand\thesubsection{\roman{subsection}} % roman numerals for subsections
\titleformat{\section}[block]{\large\scshape\centering}{\thesection.}{1em}{} % Change the look of the section titles
\titleformat{\subsection}[block]{\large}{\thesubsection.}{1em}{} % Change the look of the section titles

\usepackage{fancyhdr} % Headers and footers
\usepackage{titling} % Customizing the title section

\usepackage{hyperref} % For hyperlinks in the PDF

\makeatletter
\newcommand{\printfnsymbol}[1]{%
  \textsuperscript{\@fnsymbol{#1}}%
}
\makeatother
\pretitle{\begin{center}\Huge\bfseries} % Article title formatting
\posttitle{\end{center}} % Article title closing formatting

\title{Performance of Recommender Systems: \\
Based on Content Navigator and Collaborative Filtering} % Article title
\author{%
\textsc{Keum Gang Cha} \\[1ex] % Your name
\normalsize Nerdfactory, Plani Inc. \\ % Your institution
\normalsize \href{mailto:chagmgang@nerdfactory.ai}{chagmgang@nerdfactory.ai} % Your email address
\and % Uncomment if 2 authors are required, duplicate these 4 lines if more
\textsc{Soo-Ryeon Lee} \\[1ex] % Second author's name
\normalsize Nerdfactory, Plani Inc. \\ % Second author's institution
\normalsize \href{mailto:sooryeon@nerdfactory.ai}{sooryeon@nerdfactory.ai} % Second author's email address
\and % Uncomment if 2 authors are required, duplicate these 4 lines if more
\textsc{Jung-Woo Lee} \\[1ex] % Second author's name
\normalsize Nerdfactory, Plani Inc. \\ % Second author's institution
\normalsize \href{mailto:chris.lee.jw93@gmail.com}{chris.lee.jw93@gmail.com} % Second author's email address
\and % Uncomment if 2 authors are required, duplicate these 4 lines if more
\textsc{Seung Bin Baik} \\[1ex] % Second author's name
\normalsize Nerdfactory, Plani Inc. \\ % Second author's institution
\normalsize \href{mailto:david@nerdfactory.ai}{david@nerdfactory.ai} % Second author's email address
}

\date{\today} % Leave empty to omit a date
\renewcommand{\maketitlehookd}{%
\begin{abstract}
\noindent  % Dummy abstract text - replace \blindtext with your abstract text
In the world of big data, many people find it difficult to access the information they need quickly and accurately. In order to overcome this, research on the system that recommends information accurately to users is continuously conducted. Collaborative Filtering is one of the famous algorithms among the most used in the industry. However, collaborative filtering is difficult to use in online systems because user recommendation is highly volatile in recommendation quality and requires computation using large matrices.
To overcome this problem, this paper proposes a method similar to database queries and a clustering method (Contents Navigator) originating from a complex network.
\end{abstract}
}

\begin{document}

% Print the title
\maketitle

%----------------------------------------------------------------------------------------
%	ARTICLE CONTENTS
%----------------------------------------------------------------------------------------

\section{Introduction}

\lettrine[nindent=0em,lines=3]
The spread of IT technology has made it easier for people to access information than ever before. This also suggests that information can accumulate as much as technology spreads. This, commonly called big data, shows contradictions that make people spend more time reaching the right information. This naturally led to research on a recommender system that provides personalized, customized information. Recommender system refers that helps identify content that a particular individual may be interested in by reflecting the opinions of users[1]. Recently, machine learning methods such as deep learning were suggested and expected to have a big impact on the recommender system, but the most frequently used algorithm is Collaborative Filtering(CF), devised by Goldberg and his colleagues in 1992[2]. However, the fundamental problem has been raised that CF is difficult to operate in systems that are in the cold start stage where there is
not enough data. To solve this fundamental problem, many researchers are focusing on overcoming Cold Start[3], but are not actively introducing the service. Of course, in addition to these issues, companies dealing with big data are concerned that adopting CF, a way to explore and update the whole system's resources and the operations of services in which customers are using their own. \\
This research aims to evaluate AIVORY's recommended performance by comparing performance of CF with Content Navigator(CN), which is a recommendation algorithm for products called AIVORY that was introduced in 2018 by NerdFactory, Plani.inc.

%------------------------------------------------

\section{AIVORY and CONTENTS NAVIGATOR}

Before a full-scale test, we will first describe CN, a key recommendation engine for AIVORY. We have addressed the problem of CF and wondered how to make appropriate recommendations. And we wondered about people's interactions to create a new recommendation algorithm. And we studied how people interact and processes they do called recommendations. Exploring how people and information are connected, we have come to the complex systems networks. What we were interested in was that people and things had a certain relationship and everything was connected[5]. And it is known that there are many networks on the Internet that need to consider weights such as the transmission speed of data, the number of passengers on the air network, the level of perception in the social network, and the response rate in the metabolic network within the cell, and that these weighting networks help us to understand network characteristics quantitatively[6]. The number of papers published since 1998 with key words on complex networks is showing an explosive increase. Starting with the 1998, Small-world network, by Watts and Strogatz[7], network research show a full-fledged increase in the publication of scale-free network[8] by Albert, Jeong and Barabasi as a catalyst. Now proving its practicality in various fields, including sociology, economics, computer engineering, and biology, in addition to physical field. We focused on this point. And we designed algorithms that based on a number of other experiences, could provide customized recommendations to specific users. The recommended target was data consisting of natural language, because unlike images and videos, natural language is likely to be used in more areas of online service and is easier for us to collect data. We started the study based on the following hypotheses. \\

1. A person's recommendation stems from the person's experience. 

2. The concept of this recommendation is derived from the thoughts of others who have had similar experiences. 

3. Adjust the confidence of the recommendation according to the feedback of the person who received the recommendation. \\

The biggest drawback of the recommender system is trust issues that inevitably depend on evaluators' qualitative assessment. We were trying to solve this problem by finding the hub at the center of the network. But it was important to redefine our own hub because focusing on finding it could simply be someone who could read a lot of content. So the recommender system we'll create gave more weight to the connected links, focusing on what many people read a lot. On the other hand, we have arranged for those who are not following us but those who are leading us to read more and recommend those who are ahead of us. We also aimed at organizing a highly favored piece of writing among new ones to those who are at the forefront, reading most of the text. And to build this into a recommender system, we organized and studied algorithms that following steps. \\

1. Morphological analysis

2. Topic Modeling

3. Map users with specific preference expressions or queries to each topic

4. Create user's content consumption(or read) path with time series information

5. Extracting core users(hub) and optimal content

%------------------------------------------------

\section{MOTIVATION AND RELATED WORK}
To compare CF and recommended performance with CN, the recommended engine of AIVORY, we found a prior study that evaluated the performance of existing CF[9] and decided to approach the data of Last FM in the same way as the study that applied CF. \\
First, we decided to use the HetRec2011 dataset provided by geupplens, and after download, we checked the actual data. In the case of prior studies, the data were collected and carried out so that we could see the difference in our dataset. So we decided to do a comparative evaluation by implementing our own original CF. Last.FM dataset has a list that 1,892 users have heard of 92,800 artists. Of course, it would be better to use all the information provided by the Last.FM dataset, but considering the CN's features, we have to know the sequence of content consumed by users, so we processed what content was consumed by the users and used only the same processed data as below to evaluate it on a par with the CF. \\
In addition, both algorithms were configured to exclude users who were recommended for accurate evaluation of recommended performance. In this study, the time it takes to organize and recommend information for recommendation was measured as the study takes into account the application to the operating system. The accuracy of the recommendation is as shown in equation (1).

\begin{equation}
\Sigma_{k=1}^{l} \Sigma_{i=1}^m P_m(x_k|u_i) \times \frac{N(x_k|u_i)}{\Sigma_{c=1}N(x_c|u_i)}
    %\Sigma_{k=1}^{l} \Sigma_{i=1}^m P_m(x_k|u_i) \times \dfrac{N(x_k|u_i)}{ \Sigma_{c=1}N(x_c|u_i)}
\end{equation}

In the above expression, $I$ is the number of all contents, $m$ is the number of users, $P_m(x_k|u_i)$ is the probability of recommending $x_k$ singer to $u_i$ users, $N(x_c|u_i)$ is the number of times $u_i$'s users actually listened to $x_c$'s singer. It is assumed that the recommended contents are well recommended if they are visible in the user's pattern(future) using currently available information. If is not visible, it is assumed that the recommendation is incorrect.\\
Describing literally the above expression, the algorithm evaluates whether the user actually listened to based on the probability of recommending the singer to the user ($u_i$).
%------------------------------------------------

\section{DATA PREPROCESSING}
\subsection{Collaborative Filtering}

CF does not reflect the temporal order due to the nature of the algorithm, but it is influenced by the aspect of the content that the user consumes in evaluating the performance. In order to reflect this influence, if recommending content or things that you have already used, we will remove them from the evaluation index. We have also built a new dataset form that reflects this situation as shown in Table 1, 2.

\begin{table}[htb]
    \centering
    \caption{HetRec2011 Original Dataset}
    \begin{tabular}{c|c|c}
            \toprule
            \cmidrule(r){1-2}
            User     & Artist Name     & Date \\
            \midrule
            A & Poets of the Fall  & 2019-09-10     \\
            A & Poets of the Fall  & 2019-09-11     \\
            A & Paradise Lost  & 2019-09-12     \\
            B & Muse  & 2019-09-11     \\
            B & Paradise Lost  & 2019-09-11     \\
            B & Poets of the Fall  & 2019-09-12     \\
            C & Paradise Lost  & 2019-09-11     \\
            C & Poets of the Fall  & 2019-09-11     \\
            C & Paradise Lost  & 2019-09-12     \\
            \bottomrule
        \end{tabular}
\end{table}

\begin{table}[htb]
    \centering
    \caption{Dataset for Collaborative Filtering}
    \begin{tabular}{c|c|c|c}
    \toprule
    \cmidrule(r){1-2}
    User     & Poets of the Fall     & Paradise Lost & Muse \\
    \midrule
    A & 1  & 0 & 0     \\
    A & 1  & 0 & 0     \\
    A & 1  & 0 & 1     \\
    B & 0  & 0 & 1     \\
    B & 0  & 1 & 1     \\
    B & 1  & 1 & 1     \\
    C & 0  & 1 & 0     \\
    C & 1  & 1 & 0     \\
    C & 1  & 1 & 0     \\
    \bottomrule
        \end{tabular}
\end{table}

\subsection{Contents Navigator}

Unlike the CF, the CN is the order in which the users consumed the content. Therefore, unlike CF, the data was preprocessed to include time series information.

\begin{itemize}
\item \textit{A = [Poets of the Fall, Paradise Lost, Muse, Paradise Lost]}
\item \textit{B = [Muse, Paradise Lost, Poets of the Fall]}
\item \textit{C = [Paradise Lost, Poets of the Fall, Paradise Lost]}
\end{itemize}

Suppose a user ($ D $) who needs a recommendation has a subscription list of [Poets of the Fall, Paradise Lost], then find user, $A$, with the most similar content consumption pattern. In that case, we recommend Muse after A has been consumed. However, when evaluating the performance as described above, the recommendation is decided and the variety of recommendation decreases. That is why CN is evaluated by probabilistic calculations of not only the most frequently seen patterns, but also all cases in which they appear.

\section{RESULT}

CF requires mathematical calculation because it is based on mathematical operation with large size matrix. However, since the CN is a recommendation method using a process like Database Search Query, time consumption is smaller than CF. \\
In this paper, both algorithms are applied to recommend artist name of Last.FM to user. Comparing time consumption of both algorithms, CF and CN require 2ms and 0.2ms to recommend specific artist to user, respectively.
In addition, CN shows great accuracy variation depending on how much information is included in specific feature. The accuracy of CF according to the amount of information that feature has is showed as follows.

\begin{figure}[htb]
  \centering
  \includegraphics[width=1\linewidth]{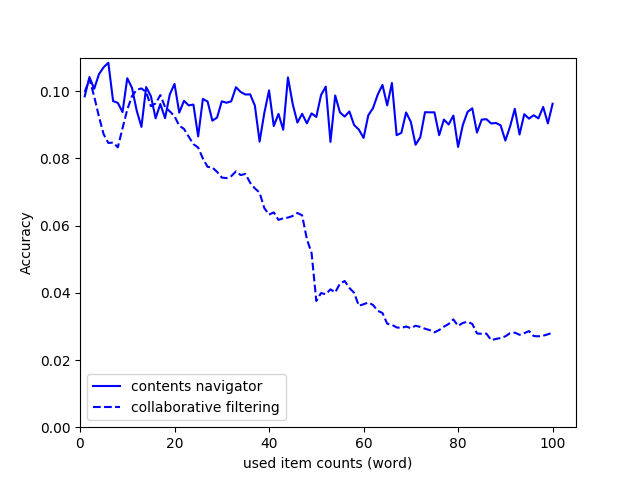}
  \label{fig:sfig1}
  \caption{Result of Performance between Collaborative Filtering and Contents Navigator}
\end{figure}

\begin{figure}[htb]
  \centering
  \includegraphics[width=1\linewidth]{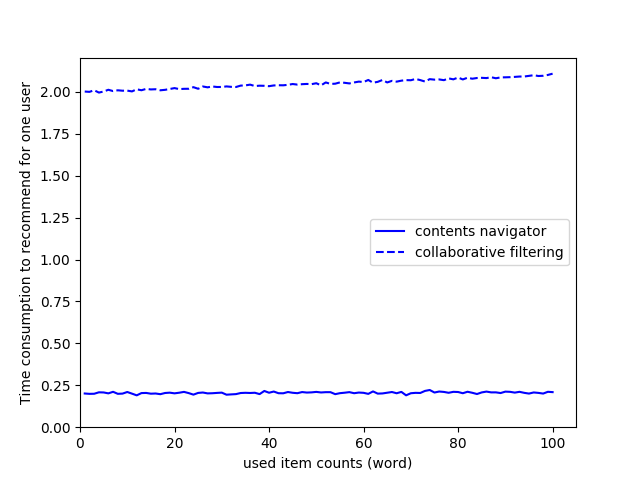}
  \label{fig:sfig1}
  \caption{Result of Performance between Collaborative Filtering and Contents Navigator}
\end{figure}

As a result, CF is highly variational algorithm whose recommendation quality is different depending on feature engineering. However, CN is using the process like DB Search Query as mentioned above. Therefore, CN has a stable recommendation quality as you can see below. \\
In other words, CF can be seen that the amount of time is consumed more than the CN when considering the size of the embedding matrix and the size of the similarity matrix. However, CN is a method of querying a kind of database rather than a matrix operation, and it is a recommendation algorithm that can adapt to an online system very quickly. In addition, the original CF recommends the content based on similarity which is based solely on mathematical techniques. However, it can be seen that recommending contents using characteristic of the complex network(people connected by contents within the web) shows higher recommendation performance.

\section{CONCLUSION}

CF measures the similarity between users using only the interaction between items and users, and then finds the preference probability of each item and recommends items. However, as shown in Sequence-to-Sequence[11] model which is frequently used in the natural language processing of deep learning, it is worth to consider the order of appearance of items, and the recommendation performance will be better when the relation characteristics of items are reflected[12]. \\
The following additional algorithms are needed to reflect the relationship between the items and the appearance order of the items. For example, Word2Vec[13] can be used to indicate that there is a relationship between adjacent tokens (contents), and Sequence-to-Sequence can be used to express the temporal characteristic of tokens (contents). However, in order to use such an algorithm in an online recommendation system, it takes a high computational power or takes a long time to operate[14].\\
This problem can be compensated by the characteristic of the user. As the user consumes the content, the subject of consumed content changes gradually. However, in a short period of time there will be no change in the theme of the content, and content that is consumed simultaneously in a short period of time may be considered to be correlated. And since users are linked to items called contents in one service, all users are connected loosely or tightly. Therefore, it is possible to make reasonable recommendation without requiring many computations by searching people who are similar to the consumption pattern of those who needs a recommendation and recommending contents that appear after that consumption pattern.

%----------------------------------------------------------------------------------------
%	REFERENCE LIST
%----------------------------------------------------------------------------------------

%----------------------------------------------------------------------------------------


\begin{thebibliography}{1}

\bibitem{kour2014real}
RESNICK, P. AND VARIAN, H. R. Recommender systems. Commun. ACM 40, 56-58. 1997.

\bibitem{kour2014real}
David Goldberg, David Nichols, Brian M. Oki, and Douglas Terry, "Using collaborative filtering to weave an information Tapestry", 1992

\bibitem{kour2014real}
Jian Wei, Jianhua He, Kai Chen, Yi Zhou AND Zuoyin Tang, "Collaborative Filtering and Deep Learning Based Recommendation System For Cold Start Items" Expert Systems with Applications , 2017

\bibitem{kour2014real}
Plani co., Ltd, System and Method for providing digital information, 1018561150000, filed March 23, 2017, and issued May 2, 2018.

\bibitem{kour2014real}
Mushon Zer-Aviv, "If everything is a network, nothing is a network", January 8, 2016.

\bibitem{kour2014real}
Barrat, M. Barthelemy, R. Pastor-Satorras, and A. Vespignani,Proc. Natl. Acad. Soc. USA 101, 3747 (2004).

\bibitem{kour2014real}
D. Watts and S. Strogatz, Nature 393, 440 (1998).

\bibitem{kour2014real}
R. Albert, H. Jeong and A.-L. Barabasi, Nature 401, 130 (1999).

\bibitem{kour2014real}
Dongjoo Lee, Sang-keun Lee, Sang-goo Lee (2009). "Considering temporal context in music recommendation based on collaborative filtering" 2009 Korea Computer Congress, Vol.3

\bibitem{kour2014real}
Stanley Milgram, "The Small-World Problem", Psychology today, 1967

\bibitem{kour2014real}
Sutskever, I., Vinyals, O., Le, Q. V. (2014). Sequence to sequence learning with neural networks. In Advances in neural information processing systems (pp. 3104-3112).

\bibitem{kour2014real}
Dong, D., Zheng, X., Zhang, R., Wang, Y. (2018, July). Recurrent Collaborative Filtering for Unifying General and Sequential Recommender. In IJCAI (pp. 3350-3356).

\bibitem{kour2014real}
Mikolov, T., Sutskever, I., Chen, K., Corrado, G. S., Dean, J. (2013). Distributed representations of words and phrases and their compositionality. In Advances in neural information processing systems (pp. 3111-3119).

\bibitem{kour2014real}
Zhang, S., Yao, L., Sun, A., Tay, Y. (2019). Deep learning based recommender system: A survey and new perspectives. ACM Computing Surveys (CSUR), 52(1), 5.D. Watts and S. Strogatz, Nature 393, 440 (1998).


\end{thebibliography}
\end{document}